\documentclass[aps,prd,twocolumn,preprintnumbers,superscriptaddress,dblfloatfix,nofootinbib]{revtex4-1}
\usepackage[utf8]{inputenc}
\usepackage{lmodern}
\usepackage{amsmath,amssymb}
\usepackage{graphicx}
\usepackage{url}
\usepackage{color}
\usepackage{enumitem}
\usepackage{subfigure}
\usepackage[dvipsnames]{xcolor}
\usepackage[colorlinks=true,breaklinks=true]{hyperref}
\hypersetup{allcolors=[rgb]{0.0 0.0 0.6},linkcolor=[rgb]{0.75 0.05 0.05}}
\usepackage{bm}
\usepackage{epsfig}
\usepackage{amsmath}
\usepackage{amssymb}
\usepackage{slashed}
\usepackage{color}
\usepackage{accents}
\usepackage{hhline}
\usepackage[dvipsnames]{xcolor}
\usepackage[colorlinks=true,breaklinks=true]{hyperref}
\hypersetup{allcolors=[rgb]{0.0 0.0 0.6},linkcolor=[rgb]{0.75 0.05 0.05}}

\renewcommand\[{\left[}

\newcommand{\exclude}[1]{}

\definecolor{offblue}{RGB}{23,80,153}

\begin{document}

\preprint{IPMU23-0053,YITP-23-169}

\title{Revisiting formation of primordial black holes in a  supercooled first-order phase transition}
	
\author{Marcos M.  Flores} 
\affiliation{Department of Physics and Astronomy, University of California, Los Angeles \\ Los Angeles, California, 90095-1547, USA}
\affiliation{
Laboratoire de Physique de l'\'{E}cole Normale Sup\'{e}rieure, ENS, Université PSL, CNRS, Sorbonne Universit\'{e}, Universit\'{e} Paris Cit\'{e}, F-75005 Paris, France
}
\author{Alexander Kusenko} 
\affiliation{Department of Physics and Astronomy, University of California, Los Angeles \\  California, 90095-1547, USA}
\affiliation{Kavli Institute for the Physics and Mathematics of the Universe (WPI), UTIAS \\The University of Tokyo, Kashiwa, Chiba 277-8583, Japan}
\author{Misao Sasaki} 
\affiliation{Kavli Institute for the Physics and Mathematics of the Universe (WPI), UTIAS \\The University of Tokyo, Kashiwa, Chiba 277-8583, Japan}
\affiliation{Center for Gravitational Physics and Quantum Information, Yukawa Institute for
Theoretical Physics,\\
Kyoto University, Kyoto 606-8502, Japan}
\affiliation{Leung Center for Cosmology and Particle Astrophysics,
National Taiwan University,\\
Taipei 10617, Taiwan}

\date{\today}
	
\begin{abstract}
We reexamine production of primordial black holes in a supercooled phase transition. While a mere overdensity associated with a surviving false-vacuum patch does not imply formation of a black hole, it is possible for such a patch to evolve and create a black hole, thanks to the gradient energy stored in the bubble wall. 

\end{abstract}

\maketitle
	

\section{Introduction}
Primordial black holes (PBHs) are an attractive candidate for dark matter~\cite{Zeldovich:1967,Hawking:1971ei,Carr:1974nx,Khlopov:1985jw,Dolgov:1992pu,Yokoyama:1995ex,Wright:1995bi,GarciaBellido:1996qt,Kawasaki:1997ju,Green:2004wb,Khlopov:2008qy,Carr:2009jm,Frampton:2010sw,Clesse:2015wea,Kawasaki:2016pql,Carr:2016drx,Inomata:2016rbd,Pi:2017gih,Inomata:2017okj,Garcia-Bellido:2017aan,Georg:2017mqk,Inomata:2017vxo,Kocsis:2017yty,Ando:2017veq,Cotner:2016cvr,Cotner:2017tir,Cotner:2018vug,Sasaki:2018dmp,Carr:2018rid,Germani:2018jgr,Banik:2018tyb,Escriva:2019phb,Germani:2019zez,1939PCPS...35..405H,Cotner:2019ykd,Kusenko:2020pcg, Flores:2020drq, deFreitasPacheco:2020wdg,Takhistov:2020vxs,DeLuca:2020agl,Dvali:2021byy,Biagetti:2021eep,Cai:2021zsp,Pi:2021dft,Carr:2021bzv,Lu:2022jnp,Harada:2022xjp,Gelmini:2023ngs,Chen:2023tzd,Chen:2023lnj,Carr:2023tpt,Domenech:2024cjn}, and also could be responsible for various astrophysical phenomena, including seeding supermassive black holes (SMBHs),~\cite{Bean:2002kx,Kawasaki:2012kn,Clesse:2015wea,Flores:2023zpf,Lu:2023xoi}  synthesis of heavy elements~\cite{Fuller:2017uyd,Takhistov:2017nmt,Takhistov:2017bpt}, explaining non-repeating fast radio bursts~\cite{Fuller:2017uyd,Abramowicz:2017zbp,Kainulainen:2021rbg}, and $G$-objects discovered in the Galactic Center~\cite{Flores:2023lll}.

A relatively slow, supercooled first-order phase transition, in which the false-vacuum ``islands" can linger, possibly creating inhomogeneities, could provide the conditions for the formation of PBHs~\cite{Sato:1981gv, Kodama:1982sf, Blau:1986cw, Liu:2021svg, Baker:2021sno, Kawana:2022olo, Gouttenoire:2023naa, Baldes:2023rqv}. We will examine the viability of this scenario, while avoiding unreliable and incorrect assumptions that have appeared in the literature. 
We will ultimately show that there is a way to form PBHs from shrinking domains of the false vacuum, thanks to the gradient energy stored in the bubble wall.  

\section{Inapplicability of a simple overdensity condition}

The physics of supercooled phase transitions is well understood. False vacuum decay, being a stochastic process, allows for the possibility that large patches of space-time remain in false vacuum, even after the percolation of bubbles in the majority of the Universe. Regions which transitioned to true vacuum subsequently reheated and have become radiation dominated. Eventually, the delayed patch may also transition into radiation. Given the difference in transition times, the radiation density in the majority of the Universe has redshifted more relative to the delayed patch. The relative difference in radiation densities generates a local overdensity.

In determining whether an overdensity of this kind might lead to formation of a PBH, it may be tempting to apply a simple overdensity criterion for black hole formation adopted from the inflationary PBH formation discussion.  However, in most cases this approach cannot succeed.  Indeed, let us define a density contrast 
\begin{equation}
\delta (x) 
:=
\frac{\rho(x) - \bar{\rho}}{\bar{\rho}}
\end{equation}

Let us suppose that some region of size $L$ with a density contrast $\delta$ remained in the false vacuum phase at a relatively late time in the transition, and it converted to the true vacuum at a time when $\bar \rho$ has decreased to a lower value than $\rho(x)$ averaged over the region $L$.  

Is there a critical value $\delta_c$ such that a patch with $\delta>\delta_c$ collapses to a black hole?  In a flat Universe, the answer is no.  Indeed, if a patch of radiation dominated Universe has a higher density, it expands in according with the Friedmann equations, at the rate that is higher for larger values of $\rho$.  

The difference with the widely discussed scenarios for PBH formation due to collapse of curvature perturbations is rooted in the fact that a curvature perturbation on some superhorizon scales makes the Universe locally nonflat. Let us briefly review the perturbation scenario and juxtapose it with the phase transition. 

In inflation-based formation mechanisms of PBHs, cosmological inflation produces a locally perturbed region with comoving size larger than the horizon size. The density contrast on this comoving slice can be related to the curvature $K$ associated with the comoving slice~\cite{Sasaki:2018dmp}.  In particular, $K$ is a function of the conserved comoving curvature perturbation, typically denoted $\mathcal{R}_c$~\cite{Lyth:2004gb}. 

Approximating $K(r)=K={\rm const.}$, one obtains the Friedmann equation: 
\begin{equation}
    H^2+\frac{K}{a^2}= \frac{8\pi G}{3}\rho,
\end{equation}
where $H=\dot a/a$, and $a$ is the scale factor.
One can now relate the density contrast to the curvature $K$. 
\begin{equation}
\Delta
:=
\frac{\rho - \bar{\rho}}{\bar{\rho}}
=
\frac{3K}{8\pi G\bar{\rho} a^2}
=
\frac{K}{H^2 a^2}, 
\end{equation}

During radiation domination, $\bar{\rho}\propto a^{-4}$, which signals that it is the curvature perturbation that induces the density perturbation.  Now the perturbed patch of the Universe evolves as a non-flat Universe with a curvature.

Once the perturbation grows to $\Delta = 1$, the expansion stops, and the patch can collapse to a black hole. Specifically, if the size of the perturbation is larger than the Jeans length, then collapse will occur. Using~\cite{Sasaki:2018dmp} 
\begin{equation}
\frac{c_s^2 k^2}{a^2} = H^2,
\end{equation}
one obtains 
\begin{equation}
1
=
\Delta(t_c)
=
\frac{K}{c_s^2 k^2},
\end{equation}
i.e., $K = c_s^2 k^2$, where $t_c$ is the time when the density perturbation $\Delta$ is one. The criterion for PBH formation is that the comoving slice density contrast re-enters the Hubble horizon with a value greater than $\delta_c = c_s^2$.
This physical understanding of the critical density criterion will be sufficient for our purposes, but much work as been done to refine the value of $\delta_c$ needed for collapse~\cite{Shibata:1999zs, Musco:2004ak, Musco:2008hv, Harada:2013epa, Harada:2023ffo}.

As discussed in the context of the inflationary scenario, all PBH formation scenarios require a clear criterion which demarcates the formation of a black hole. The simplest requirement, which we will call the Schwarzschild criterion, requires that the mass enclosed in a given volume fit within that same mass's Schwarzschild radius. This can be generalized to the so-called hoop conditions~\cite{Thorne:1972ji}, but spiritually resembles the Schwarzschild criterion so we will not differentiate between the two. On the other hand, as we have seen in the inflationary context, there are other conditions which signal the formation of a black hole. 

While the critical density criterion established in inflationary scenarios appears to be applicable generically, there are numerous subtleties which forbid its application to overdense regions which did not originate from inflationary perturbations.
The generation of large subhorizon overdensities, including those larger than the critical density criterion, do not signal the formation of a black hole. The existence of structure in our present-day Universe makes this point obvious. Naturally, subhorizon scale black holes can form but the criterion for formation should instead be replaced by the Schwarzschild criterion. 

It may also seem reasonable to apply the critical density criterion to horizon-sized density perturbations. This too proves problematic. In the inflationary picture the generation of an overdensity is linked to the existence of a curvature perturbation. It is this curvature perturbation which provides a contracting solution for the Friedmann equations and the subsequent formation of a black hole. Without this curvature component, the equations of motion describing any Hubble-sized patch containing radiation provide only expansionary solutions. Via this argument, the generation of a horizon-sized radiation overdensity is not sufficient to signal the formation of a black hole.

In this paper, we will demonstrate that PBHs can form from supercooled phase transitions. Instead of relying on the critical collapse criteria, we will instead use the Schwarzschild criterion. A false-vacuum region surrounded by true vacuum and radiation naturally implies the existence of a domain wall separating the two regions. The pressure differential between these two regions pushes the domain wall boundary inward. During this contraction, the energy density of the volume remains constant. However, the domain wall energy per unit volume scales as $1/R$. Eventually, the domain walls will fit within the Schwartzschild radius associated with the false vacuum region leading to the formation of a PBH.

In Sec.~\ref{sec:FalseVac_Potenital} we review false vacuum decay and specify the scalar potential used in our study. In Sec.~\ref{sec:BkgrndEvo} we will discuss the background evolution of the Universe and the dynamics of bubble formation and evolution. In Sec.~\ref{sec:FormationPBHs}, the domain wall equation of motion will be discussed as well as the formation of PBHs in this scenario. Section~\ref{sec:ProbabilityConsiderations} will discuss the probability of generating large supercooled regions and present a formula for the PBH abundance. Finally, in Sec.~\ref{sec:ResultsDiscussion} we will provide benchmark points of parameter space which allow for the generation of PBHs relevant for explaining all of DM or which may be consistent with microlensing observations.

\section{False vacuum decay rate}
\label{sec:FalseVac_Potenital}

False vacuum decay is a well-studied phenomena with a plethora of applications to early Universe physics~\cite{Coleman:1977py, Callan:1977pt}. The tunneling rate from the false vacuum state to the true vacuum will be the primary focus of this section. The tunneling rate determines the nucleation rate of true vacuum bubbles. Tunneling can occur either thermally, or via quantum tunneling. 
In the case of our interest the thermal tunneling rate is appropriate:
\begin{equation}
\Gamma 
=T^4\left( \frac{S_3}{2\pi T} \right)^{3/2} e^{-S_3/T},
\end{equation}
where $\Gamma$ denotes the tunneling rate per unit volume,

and $S_3$ is the three-dimensional Euclidean action.

In this work, we will assume the nucleated bubbles are spherically symmetric and produced by a single scalar field. Under these assumptions, the $3$-dimensional, $\mathrm{O}(3)$-symmetric Euclidian action is given by
\begin{equation}
\label{eq:EuclAction}
S_3 = 4\pi\int\ dr\ r^{2}
\left[\frac{1}{2}\left(\frac{d\phi}{dr}\right)^2 + V(\phi)
\right]
.
\end{equation}
Extremizing the action given by Eq.~\eqref{eq:EuclAction} requires solving the equations of motion,
\begin{equation}
\label{eq:BounceDE}
\frac{d^2\phi}{dr^2} + \left(\frac{2}{r}\right)\frac{d\phi}{dr} = \frac{\partial V}{\partial \phi}
\end{equation}
subject to the boundary condition that
\begin{equation}
\label{eq:BounceBCs}
\phi(r\to\infty) = \phi_-,
\qquad
\left.
\frac{d\phi}{dr}
\right|_{r = 0}
=
0
\end{equation}
where $\phi_-$ corresponds value of $\phi$ at the true minimum of the potential. We have also assumed that the metastable minimum lies at the origin. The solution of Eq.~\eqref{eq:BounceDE},
called a thermal instanton, may be substituted into the action to find the tunneling rate.

For our purposes, we will examine a simple scalar theory whose properties allow for strong phase transitions. In particular, we will consider a finite-temperature effective potential of the form
\begin{equation}
\label{eq:GenPotential}
V(\phi, T) = \frac{1}{2}m^2(T)\phi^2 
- \frac{\delta(T)}{3}\phi^3 + \frac{\lambda(T)}{4}\phi^4.
\end{equation}
It is useful to define the dimensionless parameter,
\begin{equation}
\kappa(T)\equiv \frac{\lambda(T)m^2(T)}{\delta^2(T)}\,,
\end{equation}
and introduce the dimensionless field and the coordinates such that $\phi\to m^2\varphi/\delta$ and $r\to \rho/m$. The potential is then re-expressed as
\begin{equation}
\label{eq:ScalarPotential}
\begin{split}
V(\varphi, T) &= \frac{m^6(T)}{\delta^2(T)}\hat{V}(\varphi, T),\\
\hat{V}(\varphi, T) &= \frac{1}{2}\varphi^2 - \frac{1}{3}\varphi^3 + \frac{\kappa(T)}{4}\varphi^4.
\end{split}
\end{equation}
Here the parameter $\kappa(T)$ lies between $-\infty < \kappa(T) < \kappa(T_c) = \kappa_c = 2/9$ where $T_c$ corresponds to the value of $T$ where the two minima of the potential are degenerate.

The action $S_3$ is written as
\begin{equation}
\begin{split}
S_3(T) &= \frac{m^{3}(T)}{\delta^2(T)}
\hat{S}_3(T),\\
\hat{S}_3(T) &= 4\pi
\int d\rho\ \rho^{2}
\left[
\frac{1}{2}\left(\frac{d\varphi}{d\rho}\right) + \hat{V}(\varphi,T)
\right].
\end{split}
\end{equation}
Determining the action for any $T$, or equivalently any $\kappa$, amounts to first numerically evaluating Eq.~\eqref{eq:BounceDE}, subject to its boundary conditions, Eq.~\eqref{eq:BounceBCs} , then evaluating the action for this solution. This is well-trodden ground, and numerous software packages have been developed which are capable of tackling precisely this problem~\cite{Masoumi:2016wot, Sato:2019wpo, Athron:2019nbd, Guada:2020xnz}.
For simplicity, we will use fits constructed in Ref.~\cite{Levi:2022bzt} for the one-parameter potential given in Eq.~\eqref{eq:ScalarPotential}. The resulting action may then be approximated by
\begin{equation}
\label{eq:FittedAction}
S_3(T)
\simeq
\frac{m^3(T)}{\delta^2(T)}
\begin{cases}
\frac{2\pi}{3(\kappa - \kappa_c)^2}\bar{B}_3(\kappa)&\kappa > 0,\\[0.25cm]
\frac{27\pi}{2}
\left(\frac{1 + \exp\left(-|\kappa|^{-1/2}\right)}{1 + |\kappa|/\kappa_c}\right)
&\kappa < 0,
\end{cases}
\end{equation}
where $\bar{B}_3(\kappa)$ is a fitting function defined in Appendix~\ref{app:BounceDetails} and which is normalized so that $\bar{B}_3(\kappa_c) = 1$. This fitting function reproduces the analytic solutions obtained in the $\kappa\to 0$ limit and also accommodates the correct $\kappa\to \kappa_c$ behavior.

To move forward, we must specify the functional form of $m(T), \delta(T)$, and $\lambda(T)$. Following Ref.~\cite{Megevand:2016lpr}, we define
\begin{equation}
\begin{split}
m^2(T) &= 2D(T^2 - T_0^2),\\[0.25cm]
\delta(T) &= 3(ET + A),\\[0.25cm]
\lambda(T)&= \lambda.
\end{split}
\label{mdelam}
\end{equation}
It is convenient to define the zero-temperature minimum $\phi_-(T = 0)\equiv v$. This enables us to express the temperature $T_0$ as
\begin{equation}
T_0^2 = 
v^2
\left(
\frac{\lambda - 3A/v}{2D}
\right)
.
\end{equation}
We can also express zero-temperature mass, $m_\phi$, as
\begin{equation}
m_\phi^2
\equiv
\frac{\partial^2 V}{\partial \phi^2}(v,0)
=
2\lambda v^2 - 3Av
.
\end{equation}
False vacuum decay can only occur once the minimum $\phi_-(T) < 0$. This occurs at the critical temperature $T_c$,
\begin{equation}
T_c = 
\frac{1}{\lambda D - E^2}
\left[
AE + 
\sqrt{
\lambda D
\left(
A^2 + (\lambda D - E^2)T_0^2
\right)}
\right]
.
\end{equation}
Additionally, the value of the field $\phi_-(T_c)\equiv \phi_c$ is given by
\begin{equation}
\phi_c
=
T_c
\left[
\frac{2(E  + A/T_c)}{\lambda}
\right]
.
\end{equation}

For the specific set of functions $\{m(T), \delta(T), \lambda(T)\}$ in \eqref{mdelam}, the parameter $\kappa$ is given by
\begin{equation}
\kappa(T) = 
\frac{2\lambda D(T^2 - T_0^2)}{9(ET + A)^2}
.
\end{equation}
Given that $\kappa$ alone determines the non-dimensional action $\hat{S}_3$, it is a simple exercise to demonstrate that the action depends only on the parameters $D$, $E$, $\lambda$ and the ratios $A/v, T/v$. The ratio $A/v$ plays a significant role in the strength of the phase transition allowed by the potential Eq.~\eqref{eq:ScalarPotential}. Therefore, for the remainder of the paper \textit{we will fix}~\cite{Megevand:2016lpr},
\begin{equation}
\label{eq:FixedParams}
E = 1/16,
\quad
D = 2.725,
\quad
\lambda=\frac{1}{8} + \frac{3}{2}\frac{A}{v}\,,
\end{equation}
where $\lambda$ is chosen so that the zero-temperature mass $m_\phi$ is given by $m_\phi=v/2$.
The requirement that the vacuum energy vanish at $T = 0$ gives
\begin{equation}
\label{eq:VacEnergyConst}
\rho_V 
= 
\frac{v^4}{4}
\left(
\lambda - \frac{2A}{v}
\right)
=
\frac{v^4}{4}
\left(
\frac{1}{8} - \frac{A}{2v}
\right)
.
\end{equation}
Note that our choice of the parameter $D$ differs from Ref.~\cite{Megevand:2016lpr} in order to ensure that a period of supercooling occurs.

Another important aspect of our scenario is the physical properties of the domain walls which will separate regions of true and false vacua. For a temperature dependent potential this calculation is non-trivial, especially considering the fact that temperature changes across the domain wall. 

We will approximate the size of the domain wall, $\ell$, and its surface energy density, $\sigma$, using the values at $T = T_c$, namely when the minima of the potential are degenerate. This is also known as the ``thin-wall regime." The solution for the potential Eq.~\eqref{eq:GenPotential} in the degenerate vacuum limit is
\begin{equation}
x =
\int_0^\phi
\frac{d\varphi}{[2V(\varphi,T_c)]^{1/2}}\,,
\end{equation}
which gives
\begin{equation}
\phi(x)
=
-
\frac{1}{2}
\phi_c
\tanh
\left(
\frac{\lambda\phi_c}{\sqrt{8}}
x
\right)
.
\end{equation}
Therefore, the characteristic scale associated with the domain wall is
\begin{equation}
\ell \sim \frac{8^{1/2}}{\lambda\phi_c}
.
\end{equation}
The surface energy density is given by
\begin{equation}
\begin{split}
\sigma
&=
\int dx
\left[
\frac{1}{2}
\left(
\frac{d\phi}{dx}
\right)^2 + V(\phi, T_c)
\right]\\[0.25cm]
&=
\int_0^{\phi_c}
[2V(\phi,T_c)]^{1/2}\ d\phi
=
\frac{\phi_c^3}{6}
\left(
\frac{\lambda}{2}
\right)^{1/2}.
\end{split}
\end{equation}
These quantities will play an important role in our discussion of the domain wall equation of motion and the conditions for the PBH formation.

\section{Background evolution and bubble dynamics}
\label{sec:BkgrndEvo}

The formation of PBHs is determined by the evolution of energy densities within different causal regions of the Universe. In the supercooled framework, there are two relevant energy components, namely radiation and the false-vacuum energy:
\begin{equation}
\rho_{\rm tot} = \rho_R + \rho_V,
\end{equation}
where generally the radiation component will include the energy density stored in relativistic bubble-wall propagation, scalar waves and plasma excitation~\cite{Gouttenoire:2023naa}. For simplicity, we will group these effects into a singular energy component $\rho_R$ and will leave a holistic study for future work.

The evolution of the scale factor, radiation energy density and false-vacuum energy density are controlled by the differential equations
\begin{align}
H^2
= 
\left(\frac{1}{a}\frac{da}{dt}\right)^2
&=
\frac{1}{3 M_{\rm Pl}^2}
\left[
\rho_{R}(t) + \rho_V(t)
\right],\label{eq:HubbleEvo}\\[0.25cm] 
\dot{\rho}_R + 4H\frac{d\rho_R}{dt} &= -\frac{d\rho_V}{dt}.\label{eq:RadEvo}
\end{align}
For the sake of evaluation, it is useful to define the dimensionless time and energy densities
\begin{equation}
\label{eq:NondimVals}
\tau \equiv \frac{H_{\rm eq}}{\sqrt{2}} t, 
\qquad
\hat{\rho}_i(\tau) = \frac{2 \rho_i(\tau)}{3M_{\rm Pl}^2 H_{\rm eq}^2}
\end{equation}
where $H_{\rm eq}$ is the Hubble parameter at equality, i.e., when $\rho_R(\tau_{\rm eq}) = \rho_V(\tau_{\rm eq})$. We will leave a detailed derivation of the non-dimensionalized differential equations in Appendix~\ref{app:NumericalApproach}. It is also useful to define a temperature, $T_{\rm eq}$, associated with the equality time, namely
\begin{equation}
H_{\rm eq}^2 = 
\frac{2\pi^2}{90}g_\star(T_{\rm eq})\frac{T_{\rm eq}^4}{M_{\rm Pl}^2}
=
\frac{2\rho_V}{3M_{\rm Pl}^2}
.
\end{equation}
From here, we see that the equality temperature is determined by $v$ and $A/v$, assuming $\lambda$ is fixed by Eq.~\eqref{eq:FixedParams}.

While the energy density $\rho_V$ is initially constant, as soon as the temperature falls below $T_c$ thermal tunneling and bubble nucleation begins to occur. Once a bubble forms at $t'$, its radius at time $t$ is given by~\cite{Turner:1992tz},
\begin{equation}
\label{eq:BubbleRadius}
R(t;t') = \int_{t'}^{t}d\tilde{t}\ \frac{v_w(\Tilde{t})}{a(\Tilde{t})}
.
\end{equation}
The expected volume of true-vacuum bubbles per unit volume of space at time $t$, $I(t)$, dictates the probability that a given point of space remains in the false vacuum state at time $t$. Given a bubble of true vacuum with coordinate radius $R(t;t')$ then~\cite{Turner:1992tz},
\begin{equation}
\label{eq:IFunction}
I(t)
=
\frac{4\pi}{3}\int_{t_c}^{t}\Gamma(t')a^3(t')R^3(t;t')\ d t'
\end{equation}
where $t_c$ corresponds to the time in which $T = T_c$ or, in other words, when phase transitions become energetically feasible. The time dependence of vacuum energy density $\rho_V(t)$ is then given by the product
\begin{equation}
\label{eq:VacEnergyTimeDep}
\rho_V(t) = \rho_{V, {\rm eq}}\exp[-I(t)],
\end{equation}
where $\rho_{V, {\rm eq}}$ is given by the constant value presented in Eq.~\eqref{eq:VacEnergyConst}.

Eqs. \eqref{eq:HubbleEvo} - \eqref{eq:VacEnergyTimeDep} describe a two-dimensional coupled, integro-differential equation. In Appendix~\ref{app:NumericalApproach} we outline how to transform this system into a set of first-order differential equations which may be evaluated with any off-the-shelf numerical methods software.

\section{Formation of PBHs}
\label{sec:FormationPBHs}

The formation of PBHs will follow the collapse of a false vacuum patch due to the pressure differential between the true and false vacuum regions. This scenario was first explored in Refs.~\cite{Sato:1981gv, Kodama:1982sf} and later refined by Ref.~\cite{Blau:1986cw}. The formalism established by Refs.~\cite{Sato:1981gv, Kodama:1982sf, Blau:1986cw} has also been utilized to describe the possible formation of PBHs from bubble nucleation during inflation~\cite{Garriga:2015fdk, Deng:2017uwc}.

Refs.~\cite{Sato:1981gv, Kodama:1982sf, Blau:1986cw} consider a circular patch of false vacuum surrounded by a region of true vacuum. This physical scenario presents two apparent paradoxes.

First, for a sufficiently large false vacuum patch, an observer deep within the false vacuum region would unambiguously observe an inflating space-time. However, an observer outside the patch would observe an unstable, collapsing region as the pressure differential between the two regions induces an inward force.

Second, if the two observers sat arbitrarily close to each other, but separated by domain wall between the two regions, each would again observe different phenomena. The inner, false vacuum observer would see the domain wall radius of curvature increasing. However, the outer true vacuum observer would not see an increase in the radius of curvature for the same reasons as discussed in the previous paragraph. Given that general relativity will ensure continuity between the two regions, the observers must agree on the observed domain-wave radius of curvature. 

The resolution to these apparent paradoxes was first identified in Ref.~\cite{Sato:1981gv} and further explored in Ref.~\cite{Blau:1986cw}. General relativity and the non-Euclidean geometry of space-time allow for a consistent solution across the de Sitter-Schwartzschild barrier. References~\cite{Sato:1981gv, Kodama:1982sf, Blau:1986cw} highlighted the possibility that bubble Universes might detach from the original space-time. Observers inside this detached region would observe an inflating Universe, while those outside would simply see a black hole. 

On the other hand, small regions of false-vacuum will never reach a state of eternal inflation.  Alternatively, these smaller false-vacuum regions may simply collapse into a ``conventional" black holes after their domain walls, and their associated energy, enters into their respective Schwarzschild radii. It is this scenario which is applicable to the formation of PBHs from supercooled phase transitions. 

The equation of motion for the domain-wall dividing the true and false vacuum regions can be derived using the Israel junction conditions~\cite{Israel:1966rt, Israel:1967zz, Blau:1986cw}. In particular, the junction conditions lead to an equation of motion for the wall radius~\cite{Blau:1986cw},
\begin{equation}
\label{eq:ConservMass}
\begin{split}
M
&=
\frac{4\pi}{3} \rho_Vr^3
- 8\pi^2 G\sigma^2 r^3\\
&\quad 
+
4\pi \sigma r^2
\left[
1 - \left(\frac{8\pi G}{3}\rho_V\right) r^2 + \dot{r}^2
\right]^{1/2}
\end{split}
\end{equation}
where $\cdot\equiv d/d\tau$ and $\tau$ is the proper time derivative as measured along the domain-wall trajectory. It is convenient to define the Hubble scales $H_V$ and $H_\sigma$ as
\begin{align}
H_V^2 &= \frac{\rho_V}{3M_{\rm Pl}^2} = \frac{H_{\rm eq}^2}{2},\\[0.25cm]
H_\sigma 
&= \frac{\sigma}{2M_{\rm Pl}^2},
\end{align}
as well as the ratio of the two,
\begin{equation}
\eta \equiv \frac{H_\sigma}{H_V}
.
\end{equation}
For the specific potential considered in Sec.~\ref{sec:FalseVac_Potenital},
\begin{equation}
\label{eq:SmallEta}
\eta 
\sim
\frac{v^3/2M_{\rm Pl}^2}
{\sqrt{
v^4/3M_{\rm Pl}^2
}}
\sim
\frac{v}{M_{\rm Pl}}\ll 1
.
\end{equation}
We will also introduce
\begin{equation}
\chi^2 \equiv
H_V^2 + H_\sigma^2
=
H_V^2
\left(
1 + \eta^2
\right)
.
\end{equation}
This allows us to define the non-dimensional quantities
\begin{equation}
z^3
\equiv
\frac{\chi^2}{2GM}r^3,
\end{equation}
and
\begin{equation}
\tau'
\equiv
\frac{\chi^2\tau}{2 H_\sigma}
.
\end{equation}
Following Ref.~\cite{Blau:1986cw} this allows us to write Eq.~\eqref{eq:ConservMass} as
\begin{equation}
\label{eq:WallEOM}
\left(
\frac{dz}{d\tau'}
\right)^2
+
U(z)
=
E,
\end{equation}
where
\begin{equation}
\begin{split}
E 
&\equiv
-\frac{4H_\sigma^2}{(2GM)^{2/3}\chi^{8/3}}\\[0.25cm]
&=
-\frac{4\eta^2}{ ( 2G M H_V )^{2/3} (1 + \eta^2)^{4/3} }
\end{split}
\end{equation}
and,
\begin{equation}
\label{eq:WallPotential}
U(z)
=
-
\left(\frac{1-z^3}{z^2}\right)^2
-
\frac{\gamma^2}{z},
\qquad
\gamma\equiv \frac{2\eta}{(1 + \eta^2)^{1/2}}.
\end{equation}
Note that, $0\leq |\gamma|\leq 2$ and that only $E$ depends on the mass $M$. Therefore, the wall equation of motion, Eq.~\eqref{eq:ConservMass}, has been reduced to the motion of a particle moving in one dimension under the influence of the potential given above.

The implicit solution to Eq.~\eqref{eq:WallEOM} is
\begin{equation}
\label{eq:WallCollapseTime}
\tau' = \int_z \frac{d z'}{\sqrt{E - U(z')}}
\end{equation}
The domain of integration is determined by the initial conditions of the bubble wall, and can lead to dramatically different physical outcomes.

Fortunately, we are only interested in the collapse time-scale and so the solution Eq.~\eqref{eq:WallCollapseTime} is sufficient. However, to establish an intuition for the process of collapse we will first examine properties of the potential, Eq.~\eqref{eq:WallPotential}. The potential, $U(z)$, is both negative and concave down for all $z$. Furthermore, the potential has one maximum, $U_m$, located at
\begin{equation}
z_m^3
=
\frac{1}{2}
\left\{
\left[
8 + 
\left(
1 - \frac{1}{2}\gamma^2
\right)^2
\right]^{1/2}
-
\left(
1 - \frac{1}{2}\gamma^2
\right)
\right\}
\end{equation}
which gives
\begin{equation}
U_m
\equiv
U(z_m)
=
-\frac{3(z_m^6 - 1)}{z_m^4}
.
\end{equation}
The maximum of the potential also allows us to define the critical mass $M_{\rm cr}$ as $E(M_{\rm cr}) = U_m$ given by
\begin{equation}
\label{eq:CritMass}
M_{\rm cr}
=
\overline{M}
\frac{\gamma^3 z_m^6\left(1 -\gamma^2/4\right)^{1/2}}{3\sqrt{3}(z_m^6 - 1)^{3/2}},
\end{equation}
where
\begin{equation}
\overline{M}
\equiv 
\frac{4\pi M_{\rm Pl}^2}{H_V}
.
\end{equation}
Observe that in the limit that $\gamma\to 0$, $M_{\rm cr}\to \bar{M}$. 

\begin{figure}[htb]
    \includegraphics[width=0.95\linewidth]{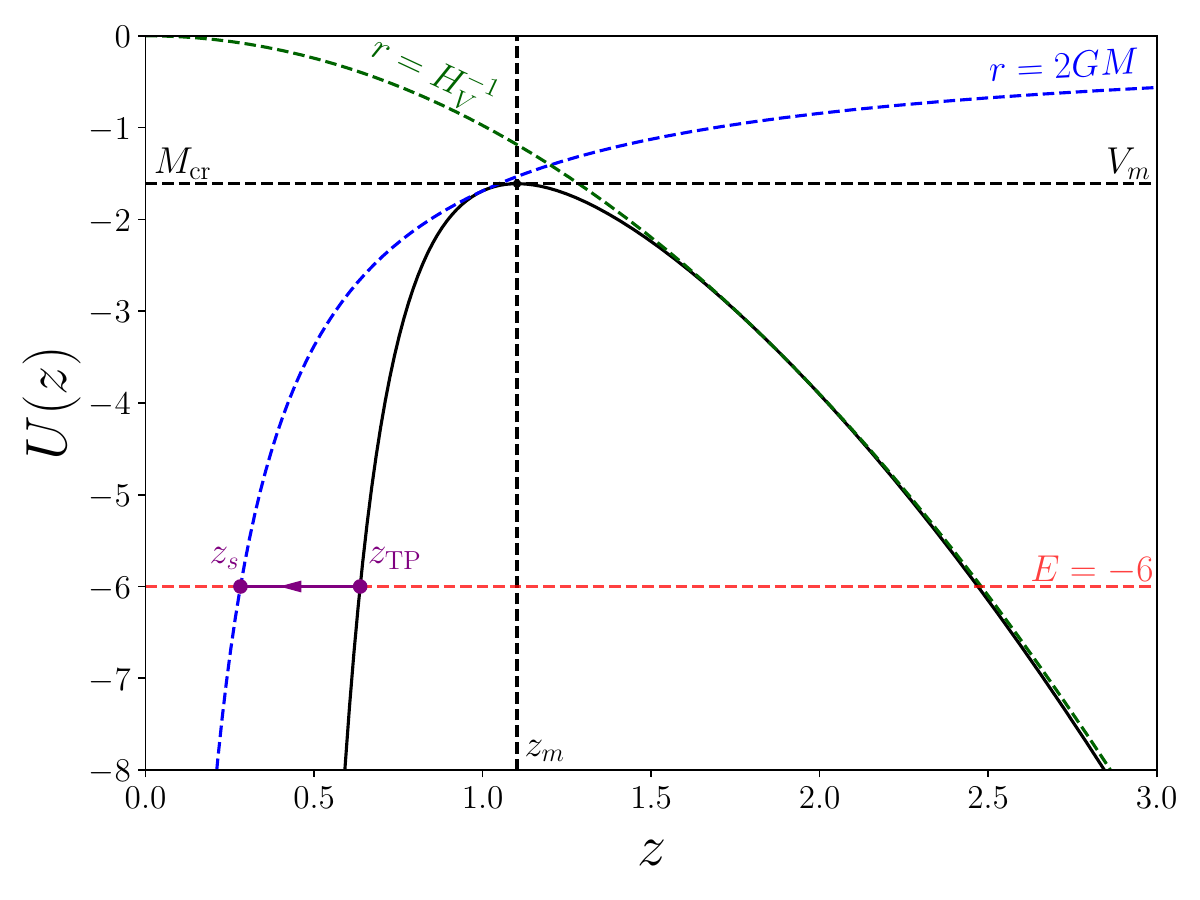}
    \caption{The nondimensional potential from Eq.~\eqref{eq:WallPotential}. The critical mass $M_{\rm cr}$ is labeled, along with the lines which specify the Schwarzschild or Hubble radius $H_V^{-1}$ for a fixed energy $E$. The purple line and arrow illustrate the motion of a wall with initial size $r_{\rm TP}$ which falls into its Schwarzschild to form a black hole. 
    For illustrative purposes we set $\gamma = 1.3$ in this figure, though $\gamma\ll1$ in realistic models. }
	\label{fig:WallPotential}
\end{figure}

Lastly, it is convenient to express the Schwarzschild radius, $r = r_s = 2GM$ and the de Sitter horizon $r = H_V^{-1}$ in terms of $z$. It is a simple exercise to show that
\begin{align}
z_s 
&=
\frac{\gamma^2}{|E|},\\
z_H
&=
\frac{\sqrt{|E|}}{\gamma(1 - \gamma^2/4)^{1/2}}
.
\end{align}
The key properties of the potential, Eq.~\eqref{eq:WallPotential}, can be observed in Fig.~\ref{fig:WallPotential}. With Eq.~\eqref{eq:SmallEta} in mind, we note that
\begin{align}
z_s(\gamma\to 0) &= (M/\bar{M})^{2/3},\\[0.25cm]
z_H(\gamma\to 0) &= (M/\bar{M})^{-1/3}
.
\end{align}
This fact, and the maximum mass defined in Eq.~\eqref{eq:CritMass}, limit the values of $z_s$ and $z_H$ in the $\gamma \to 0 $, or equivalently, the $\eta\to 0$ limit.
 
For the purposes of forming PBHs from supercooled first-order phase transitions, regions of false vacuum will originally be generated with $z > 0$  ($r > 0$) within an inward velocity, i.e., moving toward $z =  0$. For a region to collapse, its radius must lie on the left side of the potential. Otherwise, the radius of the region will simply oscillate between the two walls of the potential or expand to infinity after reflecting off of the right-most side of the potential.

We will define the radius $z_{\rm TP}$, and correspondingly $r_{\rm TP}$, as the turning point radius in which a system with fixed mass $M$, and therefore fixed ``energy" $E$, intersects the left-most wall of the potential. Based on our previous discussion, the regions which can collapse into PBHs must have initial radii between $z_s$ and $z_{\rm TP}$. As an example, the purple line in Fig.~\ref{fig:WallPotential} illustrates the path taken by a wall with $E = -6$, beginning $z_{\rm TP}$ and falling inward toward $z_s$.

Finally, in order for a PBH to form in this scenario, a number conditions must hold. The first condition is related to the collapse time $\tau'$, or alternatively its dimensionful counterpart $\tau$. We require that $\tau$ be much smaller than the Hubble time, i.e.,
\begin{equation}
\frac{\tau}{H_{V}^{-1}}
=
\frac{2\eta}{(1 + \eta^2)}\tau'
\ll 1 
.
\end{equation}
To further avoid complications during collapse we will also require that bubble nucleation rate is slow compared to the rate of collapse. Given that $\Gamma$ is the rate per volume, the corresponding nuclation time scale within a patch of radius size $z$ at time $t$ is
\begin{equation}
\tau_{\rm nuc}(z; t)
=
\left[
\frac{4\pi}{3}
r_s^3
\left(
\frac{z}{z_s}
\right)^3 \Gamma(t)
\right]^{-1}
.
\end{equation}
Therefore we will also require that,
\begin{equation}
\label{eq:NuclCheck}
\frac{\tau}{\tau_{\rm nuc}(z; t)}
\ll 
1
\end{equation}
where the values $z$ and $t$ will be specified in Sec.~\ref{sec:ProbabilityConsiderations}. 

The final PBH formation condition we will specify involves the physical size of the surrounding domain wall. In particular, we require that the size of the domain wall, $\ell$, fits within the Schwartzschild radius of a black hole with mass specified by Eq.~\eqref{eq:ConservMass}.  This leads to the condition,
\begin{equation}
\begin{split}
1
&\geq  
\frac{\ell}{2G M}\\[0.25cm]
&\simeq 
6.7\times 10^{-7}
\left(
\frac{10^{-13}\ M_\odot}{M}
\right)
\left(
\frac{1\ {\rm GeV}}{v}
\right)
\end{split}
\end{equation}
which is quite easy to satisfy. To reiterate, $\ell$ is the thin-wall approximated value of the domain wall size. However, it is difficult to believe that any correction to the size of the domain wall will exceed the six orders of magnitude needed to overcome the above inequality.

So long as the above conditions are met a PBH will form following the collapse of the false vacuum region as described by~Eq.~\eqref{eq:WallEOM}.

\section{The possibility of an island}
\label{sec:ProbabilityConsiderations}

Crucial to the supercooled phase transition paradigm is the existence of large islands of the false vacuum in which the phase transition is delayed. As is expected, the majority of bubble nucleation will occur when
\begin{equation}
\Gamma(t_n) \simeq H^4(t_n)
\end{equation}
where $t_n$, and the corresponding temperature $T_n$, satisfies the above equality. In our realisation of PBH formation from supercooled phase transitions, the probability of collapse corresponds to the probability that a patch with comoving radius $\mathcal{R}$ can exist until some late time. Specifically,
\begin{equation}
\mathcal{P}_{\rm coll}
\equiv 
\mathcal{P}_{\rm surv}(t_{n_i};\mathcal{R})
\end{equation}
where $\mathcal{P}_{\rm surv}(t_{n_i};\mathcal{R})$ specifies the probability that a patch with comoving radius $\mathcal{R}$ has survived up to time $t_{n_i}$. As one expects, $t_n < t_{n_i}$.

The collapse probability can be written explicitly as~\cite{Sato:1981gv, Kodama:1982sf}
\begin{equation}
\label{eq:FullProb}
\ln\mathcal{P}_{\rm surv}(t_{n_i};\mathcal{R})
=
-
\int_{t_c}^{t_{n_i}}dt'\ \Gamma(t')a^3(t')\mathcal{V}(t'; t_{n_i})
\end{equation}
where
\begin{equation}
\mathcal{V}(t'; t_{n_i}) 
= 
\frac{4\pi}{3}
\left[
\mathcal{R}(t_{n_i})
+
R(t_{n_i};t')
\right]^3
\end{equation}
is the volume corresponding to the space-like slice at $t'$ of a patch with comoving radius $\mathcal{R}$ at time $t_{n_i}$. It is convenient to define the ratio
\begin{equation}
\xi = \frac{\mathcal{R}(t_{n_i}) }{[a(t_{n_i})H(t_{n_i})]^{-1}}
\end{equation}
which allows us to directly compare the size of the comoving radius $\mathcal{R}$ and the comoving Hubble radius at time $t_{n_i}$. In particular, for $\xi < 1$ we are dealing with subhorizon patches, while for $\xi > 1$ the delayed patch is larger than the horizon size. For the parameters relevant for our discussion, we will always be well within the subhorizon case.

\begin{figure}[htb]
    \includegraphics[width=0.95\linewidth]{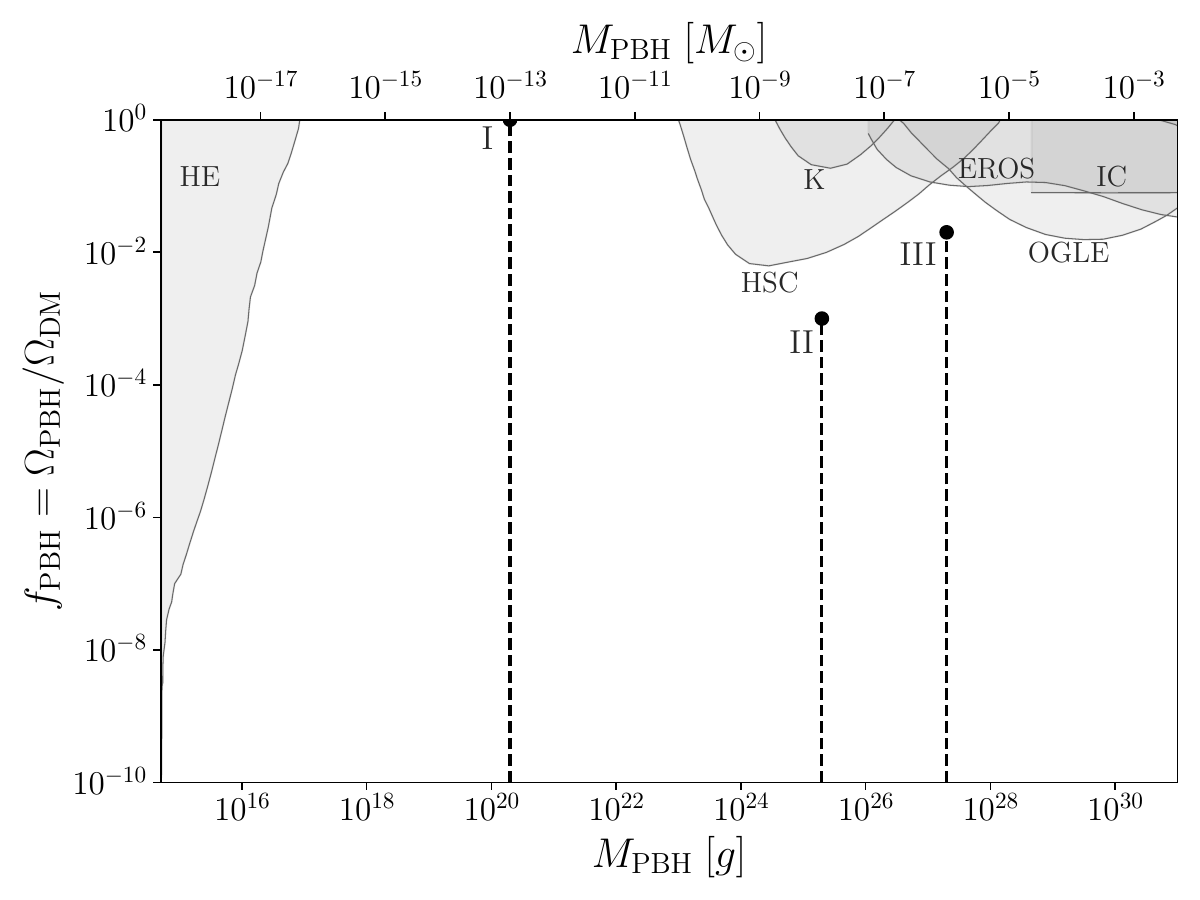}
    \caption{PBH masses generated from our realization of the supercooled phase transition scenario compared with observational constraints. A summary of the constraints can be found in Refs.~\cite{Carr:2020gox, Carr:2020xqk}. Here, HE specifies Hawking evaporation constraints. The remaining constraints are due to microlensing observations by the Hyper Suprime-Cam (HSC)~\cite{Niikura:2017zjd, Smyth:2019whb}, Kepler (K)~\cite{Griest:2013esa, Griest:2013aaa}, EROS~\cite{EROS-2:2006ryy}, OGLE~\cite{Niikura:2019kqi} and Icarus (IC)~\cite{Oguri:2017ock}.}
	\label{fig:PBHConstraints}
\end{figure} 

As discussed in Sec.~\ref{sec:FormationPBHs}, the initial radius of the false vacuum region must lie within $z_s$ and $z_{\rm TP}$ for a ``conventional" black hole to form. This limits the acceptable values of $\xi$ in this circumstance. For simplicity, we will perform our calculations for regions size $z_{\rm TP}$ at $t_{n_i}$. It is these values of $z$ and $t$ which will be used in the nucleation rate condition specified by Eq.~\eqref{eq:NuclCheck}.

Calculating the PBH abundance from supercooled phase transitions is non-trivial. In general one has to consider the collapse of false vacuum patches with varying sizes. This will lead to a non-monochromatic distribution of PBHs. As is standard, we express the present-day PBH abundance as
\begin{equation}
f_{\rm PBH}
\equiv
\frac{\rho_{\rm PBH, 0}}{\rho_{\rm DM, 0}}.
\end{equation}
In this particular context,
\begin{equation}
\begin{split}
f_{\rm PBH}
&=
\frac{1}{\rho_{\rm DM, 0}}
\left(
\frac{4\pi}{3}H_0^{-3}
\right)^{-1}\\[0.25cm]
&\times
\int
M
\mathcal{N}_{\rm patches}(M)
\mathcal{P}_{\rm surv}(M)
g(M)\  dM.
\end{split}
\end{equation}
Here, $\mathcal{N}_{\rm patches}$ denotes the number of patches within the present day horizon which have the associated mass $M$. Eq.~\eqref{eq:ConservMass} provides the relation between the mass $M$ and the initial size of the patch. The function $g(M)$ describes the distribution of patches. In the absence of an analytic expression for $g(M)$, we will utilize the monochromatic limit
\begin{equation}
g(M) = \delta(M - M_{\rm PBH})  
\end{equation}
which will approximate the peak of the true distribution. This leads to
\begin{equation}
f_{\rm PBH}
=
\mathcal{P}_{\rm surv}\frac{M_{\rm PBH}\mathcal{N}_{\rm patches}}
{\rho_{\rm DM, 0}\frac{4\pi}{3}H_0^{-3}}.
\end{equation}
Here, 
\begin{equation}
\mathcal{N}_{\rm patches}
\simeq
\frac{1}{\xi^3}
\cdot
\left(
\frac{a_{\rm eq}H_{\rm eq}}{a_0H_0}
\right)^3
\end{equation}
where we used the fact that
\begin{equation}
a(t_{n_i})H(t_{n_i})\simeq a(T_{\rm eq})H(T_{\rm eq})
\end{equation}
within the delayed, inflating patch. The expression for $\mathcal{N}_{\rm patches}$ can be further simplified by utilizing conservation of entropy, namely
\begin{equation}
g_{*S}(T_{\rm eq})a_{\rm eq}^3T_{\rm eq}^3 
= 
g_{*S}(T_{0})a_{0}^3T_{0}^3
.
\end{equation}
This allows us to express $\mathcal{N}_{\rm patches}$ as
\begin{equation}
\mathcal{N}_{\rm patches}
\simeq
\frac{1}{\xi^3}
\cdot
\frac{g_{*S}(T_0)}{g_{*S}(T_{\rm eq})}
\left(
\frac{T_0 H_{\rm eq}}{T_{\rm eq} H_0}
\right)^3
\end{equation}
where we used $g_{*S}(T_0) = 3.94$, $T_0 = 2.7\ {\rm K} = 2.4\times 10^{-13}$ GeV and $H_0 = 1.4\times 10^{-42}$ GeV.

This leads to a parametrization of $f_{\rm PBH}$ of the form
\begin{equation}
\begin{split}
\label{eq:fPBHParam}
f_{\rm PBH}
&\simeq
\left(
\frac{7.3\times 10^{-5}}{\xi}
\right)^3
\left(
\frac{\mathcal{P}_{\rm surv}}{3\times 10^{-8}}
\right)
\times \\[0.25cm]
&\quad
\times
\left(
\frac{M_{\rm PBH}}{10^{-13}\ M_{\odot}}
\right)
\left(
\frac{g_*(T_{\rm eq})}{100}
\right)^{1/2}
\left(
\frac{T_{\rm eq}}{150\ {\rm MeV}}
\right)^3
\end{split}
\end{equation}

\section{Results \& Discussion}
\label{sec:ResultsDiscussion}
The above formalism enables us to identify numerous points in parameter space which could have important phenomenological implications. For the scope of this work, we will highlight a few points of interest and leave an exhaustive parameter search for later work. Table~\ref{tab:PBHParams} presents three sets of parameters which offer PBH populations relevant for dark matter, microlensing observations and gravitational wave detectors. A summary of our results can also be seen in Fig.~\ref{fig:PBHConstraints} alongside the present-day landscape of observational constraints for PBHs.

\begin{table}[htb]
    \centering
    \begin{tabular}{c|c|c|c|c|c}
       Model & $v$ [GeV] & $A/v$  & $M_{\rm PBH}$\ [$M_\odot$] & $t_{n_i}/t_{\rm eq}$ & $f_{\rm PBH}$\\
       \hhline{======}
       I     & 1 & 0.1052 & $10^{-13}$ & 1.4106 & 1\\
       II & 10 & 0.1040  & $10^{-8}$  & 1.3305 & $10^{-3}$ \\
       III & 50 & 0.1037 & $10^{-6}$  & 1.7332 & $2\times 10^{-2}$\\
        \hhline{======}
    \end{tabular}
    \caption{Parameters which generate PBHs with masses relevant for dark matter or microlensing experiments. }
    \label{tab:PBHParams}
\end{table}

Given that our scenario relies on the existence of a false vacuum region surrounded by primarily by true vacuum we require that the ratio $\rho_{V}^{\rm bkg}/\rho_{\rm rad}^{\rm bkg}\sim \mathcal{O}(0.1)$ at $t = t_{n_i}$. 

\begin{figure}[htb]
    \includegraphics[width=0.95\linewidth]{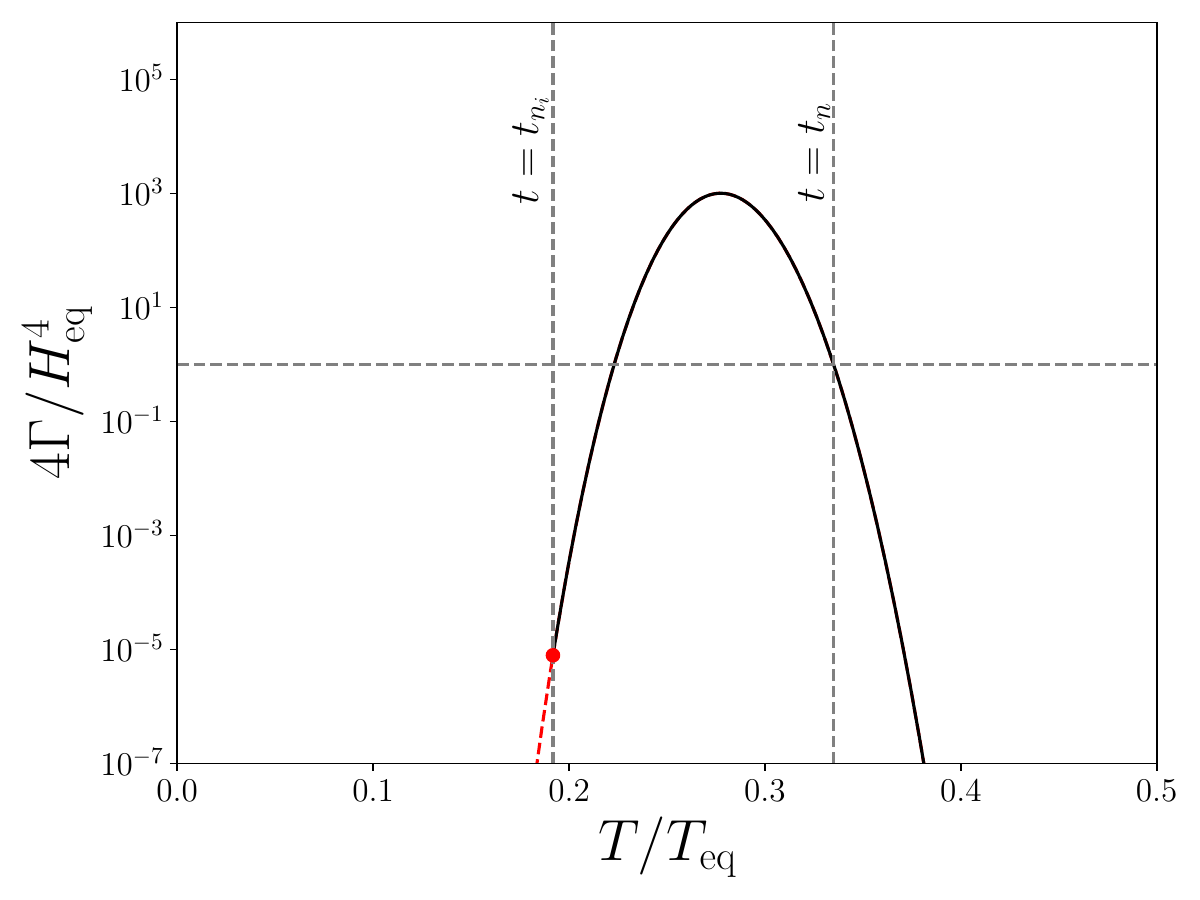}
    \caption{The nucleation rate per unit volume normalized according to the nondimensionalization specified in \eqref{eq:NonDimRate}. The maximum value of the rate is relatively small which allows for the majority of the Universe to transition to radiation domination before the false vacuum island collapses into a black hole. The parameters utilized correspond to Model I in Table~\ref{tab:PBHParams}.}
	\label{fig:NucRate}
\end{figure}

Figure~\ref{fig:NucRate} shows the nucleation rate associated with $M_{\rm PBH} = 10^{-13}\ M_\odot$. A notable feature of the decay rate is that $(\Gamma/H^4)_{\max} \lesssim 10^4$. In our parameter search, this feature has generally been required for the successful formation of PBHs. Adjusting the ratio $A/v$ can lead to extremely large values of $(\Gamma/H^4)_{\max}$. After equaling one, the ratio $\Gamma/H^4$ increases extremely rapidly. Since the nucleation rate is so large, the probability to survive any appreciable time after $t_{n}$ immediately becomes negligible (see Eq.~\eqref{eq:FullProb}).
Given that the delay times needed for PBH formation occur significantly after $t_{n}$, we also note that the approximation
\begin{equation}
\Gamma(t)
\sim
\Gamma_{n}
e^{-\beta(t - t_n)}
\end{equation}
where
\begin{equation}
\beta \equiv
\left.
H T
\frac{dS_3}{dT}
\right|_{T = T_n}
\end{equation}
\textit{does not hold} in this scenario.

In summary, we've presented the possibility that supercooled phase transitions can lead to the formation of PBHs. Large islands of false vacuum have the possibility to persist after a short intermediate de Sitter phase. The domain waves separating the false and true vacuum regions is driven inward due to the pressure differential between the two phases. These domain waves fall within the Schwarzschild radius of the false vacuum region, leading to the formation of a black hole. This framework is significantly different from existing literature which instead relied on the critical density condition to signal the formation of PBHs~\cite{Liu:2021svg, Baker:2021sno, Kawana:2022olo, Gouttenoire:2023naa, Gouttenoire:2023pxh, Baldes:2023rqv}. 
We will also point to existing literature, such as Refs.~\cite{Jinno:2023vnr, Lewicki:2023ioy, Lewicki:2024ghw}, which also previously examined the formation of PBHs from super-cooled phase transitions, but which have instead opted to use the Schwarzschild formation condition.

While not emphasized here, it is well known that first-order phase transitions can lead to sizeable gravitational wave signals due to collision of bubble walls. The amplitude for such a signal is determined by the ratio~\cite{Konstandin:2017sat, Cutting:2018tjt, Cutting:2020nla, Lewicki:2020jiv}
\begin{equation}
\Omega_{\rm GW}
\sim 10^{-6}
\left(
\frac{H}{\beta}
\right)^2
\end{equation}
while the peak frequency is given by
\begin{equation}
f_{\rm peak}
\sim 
1\ {\rm mHz}
\left(
\frac{\beta}{H}
\right)
\left(
\frac{T_{\rm RH}}{100\ {\rm TeV}}
\right)
\end{equation}
where $T_{\rm RH}$ is the temperature when the majority of the Universe is restored to a radiation dominated era.

For the parameters we've explored, $\beta/H \sim 100$ and $T_{\rm RH}\lesssim 0.1$ GeV leading to $\Omega_{\rm GW} \sim 10^{-10}$ with $f_{\rm peak}\sim 10^{-7}$ Hz. Frequency is relevant for future pulsar timing arrays measurements. That being said, it would be difficult to associate any gravitational waves in this frequency range directly with the formation of PBHs. However, such a gravitational wave signal might help in distinguishing this PBH formation mechanism from the expected signals of others.

\begin{acknowledgments}
We thank K.~Petraki for helpful discussions. 
This work was supported by the European Union’s Horizon 2020 research and innovation programme under grant agreement No 101002846, ERC CoG CosmoChart. The work of A.K. was supported by the U.S. Department of Energy (DOE)
Grant No. DE-SC0009937. The work of A.K. and M. S. was also supported by World Premier International Research 
Center Initiative (WPI), MEXT, Japan, and by Japan Society for the Promotion of Science (JSPS) KAKENHI Grant No. JP20H05853.

\end{acknowledgments}


\bibliography{biblio}


\appendix

\section{Bounce Details}
\label{app:BounceDetails}

Calculating the single parameter bounce action, as defined by Eqs.~\eqref{eq:EuclAction} - \eqref{eq:FittedAction}, is a well understood procedure. As discussed in the main text, a potential of the form of Eq.~\eqref{eq:GenPotential} can be redefined through field and coordinate definitions to obtain an equation of motion which depends only on one, temperature dependent parameter $\kappa$.  In order to obtain the bounce solution, one must then solve Eq.~\eqref{eq:BounceDE} subject to the boundary conditions given in Eq.~\eqref{eq:BounceBCs}. Generally, this must be done numerically and many software packages have been written which can solve this simple bounce problem, and it's generalizations~\cite{Masoumi:2016wot, Sato:2019wpo, Athron:2019nbd, Guada:2020xnz}.

For simplicity, we will utilize a fitting function for the $\kappa$ dependent bounce action $S_3$ obtained in Ref.~\cite{Levi:2022bzt}. In particular for, $\kappa > 0$

\begin{equation}
S_3(T)
\simeq
\frac{m^3(T)}{\delta^2(T)}
\frac{2\pi}{3(\kappa - \kappa_c)^2}\bar{B}_3(\kappa) 
\end{equation}
where $\bar{B}_3(\kappa)$ is obtained from numerical fits and given by~\cite{Levi:2022bzt},
\begin{widetext}
\begin{equation}
\bar{B}_3(\kappa)
=
\frac{16}{243}
\left[
1 - 38.23\left(\kappa - \frac{2}{9}\right)
+ 115.26\left(\kappa - \frac{2}{9}\right)^2 
+ 58.07\sqrt{\kappa}\left(\kappa - \frac{2}{9}\right)^2
+ 229.07\kappa\left(\kappa - \frac{2}{9}\right)^2
\right]
.
\end{equation}
\end{widetext}

We verified that this fitting function is in agreement the results obtained in Ref.~\cite{Megevand:2016lpr} as well as with our own numerical implementation of the the one-parameter bounce solution.

\section{Numerical Approach}
\label{app:NumericalApproach}

In this section, we hope to illustrate one possible methodology for solving \eqref{eq:HubbleEvo} - \eqref{eq:VacEnergyTimeDep}. Our first step is rewriting the background evolution equations Eqs. \eqref{eq:HubbleEvo}, \eqref{eq:RadEvo} in terms of the dimensionless time variable $\tau$ (see Eq.~\ref{eq:NondimVals}). This simple exercise yields,
\begin{align}
a'(\tau)
&=
a(\tau)
\left[
\hat{\rho}_R(\tau) + \hat{\rho}_V(\tau)
\right]^{1/2}\\[0.25cm]
\hat{\rho}_R'(\tau)
&+
4 \frac{a'(\tau)}{a(\tau)} \hat{\rho}_R = -\hat{\rho}_V'(\tau)
\end{align}
where $\prime \equiv d/d\tau$. We will normalize the scale factor such that $a(\tau_{\rm eq}) = 1$. In addition, we must also non-dimensionalize Eqs.~\eqref{eq:BubbleRadius} and \eqref{eq:IFunction}. It is a simple exercise to show that this leads to
\begin{align}
\hat{R}(\tau;\tau') &= \int_{\tau'}^{\tau}d\Tilde{\tau}\ \frac{v_w(\Tilde{\tau})}{a(\Tilde{\tau})}\\[0.25cm]
I(\tau)
&=
\frac{4\pi}{3}
\int_{\tau_c}^{\tau}
\hat{\Gamma}(\tau')a^3(\tau')\hat{R}^3(\tau;\tau')\ d\tau' \label{eq:NonDimRate}
.
\end{align}
The nondimensional radius $\hat{R}$ and rate $\hat{\Gamma}$ are related to their dimensionful quantities by
\begin{align}
R(\tau;\tau') &= \frac{\sqrt{2}}{H_{\rm eq}}\hat{R}(\tau;\tau')\\
\Gamma(\tau)  &= \frac{H_{\rm eq}^4}{4}\hat{\Gamma}(\tau)
.
\end{align}
The next step in our method will be to define,
\begin{equation}
r(\tau)\equiv
\int_{\tau_c}^{\tau}d\Tilde{\tau}\ \frac{v_w(\Tilde{\tau})}{a(\Tilde{\tau})}
\end{equation}
such that
\begin{equation}
\hat{R}(\tau;\tau') = r(\tau) - r(\tau')
\end{equation}
and
\begin{equation}
r'(\tau) = \frac{v_w(\tau)}{a(\tau)},
\quad
r(\tau \leq \tau_c) = 0
.
\end{equation}
Furthermore, we can define
\begin{equation}
v_i(\tau)
=
\int_{\tau_c}^{\tau}
\hat{\Gamma}(\tau')a^3(\tau')r^i(\tau')\ d\tau'
\end{equation}
such that
\begin{equation}
v_i'(\tau)
=
\hat{\Gamma}(\tau)a^3(\tau)r^i(\tau),
\quad
v_i(\tau \leq \tau_c) = 0
.
\end{equation}
This enables us to write $I(\tau)$ as
\begin{equation}
\begin{split}
I(\tau)
=
\frac{4\pi}{3}
\left[
r^3(\tau)v_0(\tau)
-
3r^2(\tau)v_1(\tau) + \right.\\
\left.
+
3r(\tau)v_2(\tau)
-
v_3(\tau)
\right].
\end{split}
\end{equation}
Thus, we have extended our two-dimensional system of integro-differential equation to a seven-dimensional system of first order ODEs,
\begin{equation}
\left\{a(\tau),\rho_R(\tau)\right\}
\longrightarrow
\left\{a(\tau),\rho_R(\tau), r(\tau), v_i(\tau)\right\}
.
\end{equation}
The initial conditions are,
\begin{equation}
\left.
\left\{a(\tau),\rho_R(\tau), r(\tau), v_i(\tau)\right\}
\right|_{\tau = \tau_c}
=
\left\{1, 1, 0, 0\right\}
.
\end{equation}
The cost of solving seven-dimensional first order ODEs is minimal, and so, this method provides quick, easy to implement strategy to examine background evolution of a supercooled phase-transition.

\end{document}